\documentclass[aps,prl,twocolumn,superscriptaddress]{revtex4-2}

\usepackage{graphicx}
\usepackage{amssymb}

\newcommand{\SIQSE}{\affiliation{1}{Shenzhen Institute for Quantum Science and Engineering, Southern University of Science and Technology, Shenzhen, Guangdong, China}}
\newcommand{\NXU}{\affiliation{2}{
School of Physics, Ningxia University, Yinchuan, 750021, China}}
\newcommand{\PKU}{\affiliation{3}{
School of Physics, Peking University, Beijing 100871, China}}

\newcommand{\IQA}{\affiliation{5}{International Quantum Academy, Shenzhen, Guangdong, China}}
\newcommand{\GDKL}{\affiliation{6}{Guangdong Provincial Key Laboratory of Quantum Science and Engineering, Southern University of Science and Technology, Shenzhen, Guangdong, China}}
\newcommand{\HFNL}{\affiliation{7}{
Shenzhen Branch, Hefei National Laboratory, Shenzhen 518048, China}}

\begin{document}


\title{Improving Quantum Optimization to Achieve Quadratic Time Complexity}

\author{Ji Jiang}
\thanks{Ji Jiang and Peisheng Huang contributed equally to this work.}
\email[Contact author: ]{jiangj3@sustech.edu.cn}
\affiliation{\SIQSE}\affiliation{\IQA}\affiliation{\GDKL}

\author{Peisheng Huang}
\thanks{Ji Jiang and Peisheng Huang contributed equally to this work.}
\affiliation{\NXU}\affiliation{\IQA}

\author{Zhiyi Wu}
\affiliation{\PKU}\affiliation{\IQA}
\author{Xuandong Sun}
\affiliation{\SIQSE}\affiliation{\IQA}\affiliation{\GDKL}
\author{Zechen Guo}
\affiliation{\SIQSE}\affiliation{\IQA}\affiliation{\GDKL}
\author{Wenhui Huang}
\affiliation{\SIQSE}\affiliation{\IQA}\affiliation{\GDKL}
\author{Libo Zhang}
\affiliation{\SIQSE}\affiliation{\IQA}\affiliation{\GDKL}
\author{Yuxuan Zhou}
\affiliation{\IQA}
\author{Jiawei Zhang}
\affiliation{\SIQSE}\affiliation{\IQA}\affiliation{\GDKL}
\author{Weijie Guo}
\affiliation{\IQA}
\author{Xiayu Linpeng}
\affiliation{\IQA}
\author{Song Liu}
\affiliation{\SIQSE}\affiliation{\IQA}\affiliation{\GDKL}\affiliation{\HFNL}
\author{Wenhui Ren}
\affiliation{\IQA}
\author{Ziyu Tao}
\affiliation{\IQA}
\author{Ji Chu}
\affiliation{\IQA}
\author{Jingjing Niu}
\affiliation{\IQA}\affiliation{\HFNL}
\author{Youpeng Zhong}
\affiliation{\SIQSE}\affiliation{\IQA}\affiliation{\GDKL}\affiliation{\HFNL}
\author{Dapeng Yu}
\affiliation{\SIQSE}\affiliation{\IQA}\affiliation{\GDKL}\affiliation{\HFNL}

\date{\today}

\begin{abstract}
Quantum Approximate Optimization Algorithm (QAOA) is a promising candidate for achieving quantum advantage in combinatorial optimization. However, its variational framework presents a long-standing challenge in selecting circuit parameters. In this work, we prove that the energy expectation produced by QAOA can be expressed as a trigonometric function of the final-level mixer parameter. Leveraging this insight, we introduce Penta-O, a level-wise parameter-setting strategy that eliminates the classical outer loop, maintains minimal sampling overhead, and ensures non-decreasing performance. This method is broadly applicable to the generic quadratic unconstrained binary optimization formulated as the Ising model. For a $p$-level QAOA, Penta-O achieves an unprecedented quadratic time complexity of $\mathcal{O}(p^2)$ and a sampling overhead proportional to $5p+1$. Through experiments and simulations, we demonstrate that QAOA enhanced by Penta-O achieves near-optimal performance with exceptional circuit depth efficiency. Our work provides a versatile tool for advancing variational quantum algorithms.
\end{abstract}

\maketitle

\textit{Introduction.} --- Combinatorial optimization has broad and highly valuable applications across various industrial and scientific fields but is often exponentially difficult to solve. An exceptional variety of these problems can be formulated as Quadratic Unconstrained Binary Optimization (QUBO)~\cite{lucas2014ising,glover2022quantum}. Due to its close connection to spin models, the QUBO model has emerged as an underpinning of the quantum computing area, positioning it as one of the key problems where quantum advantage might be realized. Notable attempts include portfolio optimization~\cite{herman2023constrained,buonaiuto2023best,PhysRevResearch.4.043204,brandhofer2022benchmarking}, flight assignment and scheduling~\cite{wang2020flight,PhysRevApplied.14.034009,PhysRevApplied.20.064025}, drug discovery~\cite{PhysRevApplied.21.034036,muller2022rna,fox2022rna,PhysRevApplied.20.014024,robert2021resource}, and solving spin glass models~\cite{kim2023evidence,pagano2020quantum,dupont2023quantum,pelofske2024scaling,PhysRevApplied.22.044074} or graph problems~\cite{harrigan2021quantum,ebadi2022quantum,PhysRevA.101.012320,PhysRevResearch.6.023294}. Most of these approaches are grounded in the variational framework, specifically the Quantum Approximate Optimization Algorithm or the Quantum Alternating Operator Ansatz (QAOA)~\cite{farhi2014quantum,hadfield2019quantum}. These approaches employ parametric quantum circuits to approximate the optimal solution, encoded as the ground state of a problem-specific Hamiltonian. While it is widely anticipated that quantum advantage will materialize at a certain scale~\cite{Farhi2022quantumapproximate,shaydulin2024evidence}, several practical challenges remain unresolved.

One of the central challenges lies in selecting circuit parameters, which are critical to algorithmic performance. Conventional methods tackle this issue by treating the quantum computer's operational loop as a black-box function, and optimizing circuit parameters using gradient descent or machine learning~\cite{PhysRevX.10.021067,PhysRevA.107.032407,blekos2024review,PhysRevResearch.2.043246,PhysRevA.106.062416}. This process, commonly referred to as the ``classical outer loop"~\cite{harrigan2021quantum}, necessitates repeated execution of quantum circuits and substantially increases the time complexity. In the worst-case scenario, the parameter optimization process itself is \textit{NP}-hard~\cite{PhysRevLett.127.120502}. Based on the parameter concentration conjecture~\cite{Farhi2022quantumapproximate,PhysRevA.104.L010401,PhysRevA.104.052419}, some heuristic methods have been developed~\cite{shaydulin2023parameter,galda2021transferability,Sureshbabu2024parametersettingin}. However, these methods cannot guarantee reliable performance across diverse problems. More recently, methods inspired by quantum Lyapunov control have proposed an innovative framework to eliminate the outer loop~\cite{PhysRevLett.129.250502,PhysRevA.106.062414,PhysRevResearch.6.033336,PhysRevResearch.6.043068,chandarana2024lyapunov,PhysRevA.109.062603,brady2024focqs}. While promising, these approaches introduce additional sampling overhead and often demand unrealistically deep circuits to achieve convergence, even for small-scale problems.

In this work, we propose \textbf{Penta-O}, a level-wise parameter-setting strategy for QAOA tailored to general QUBO instances. By leveraging analytical insights into the dependence between the energy expectation and circuit parameters, our method entirely circumvents the classical outer loop while ensuring non-decreasing performance with exceptional circuit depth. Notably, for a $p$-level QAOA, Penta-O achieves an unprecedented quadratic-time complexity of $\mathcal{O}(p^2)$ and a sampling overhead proportional to $5p+1$. Benchmark results demonstrate that Penta-O enables QAOA to achieve near-optimal performance across diverse instances. Remarkably, it matches the performance of existing methods with comparable sampling overhead while requiring only one-tenth of their circuit depth.

\begin{figure*}
\includegraphics[width=1.0\textwidth]{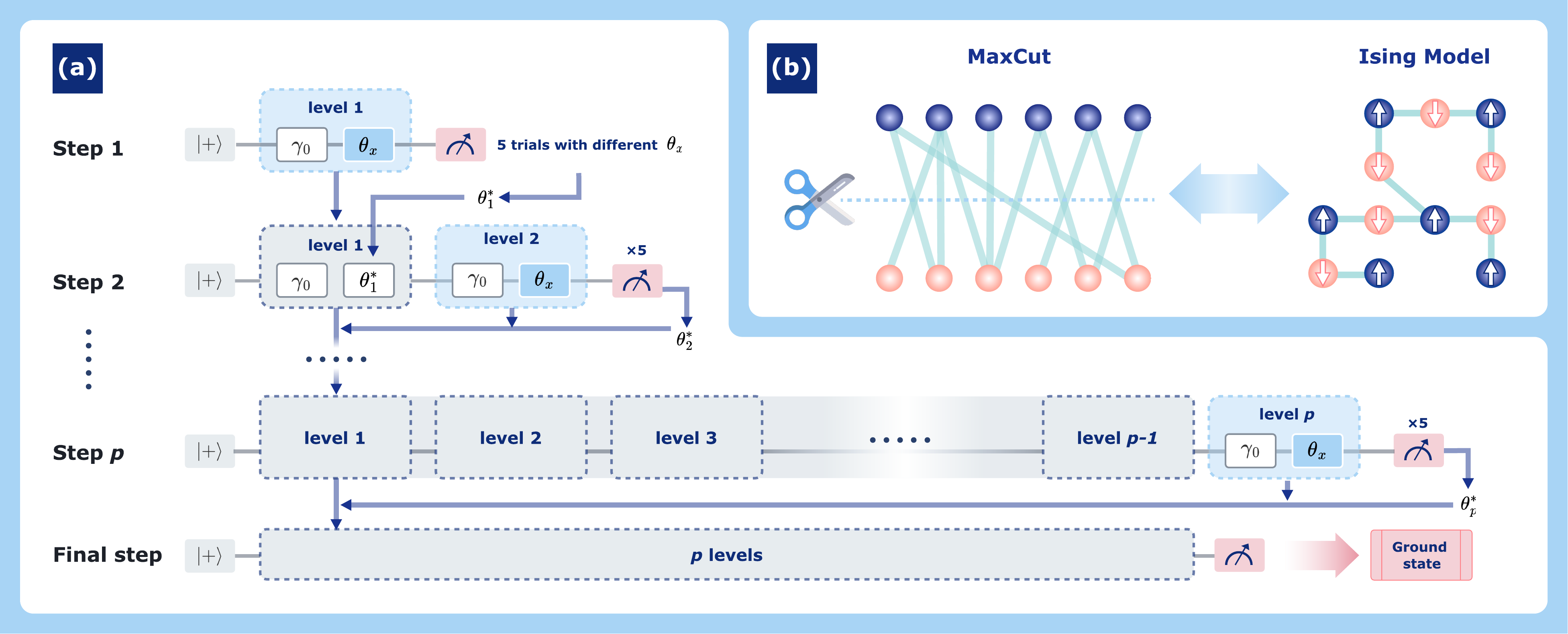}
\caption{\label{fig:overview} (a) Visualization of Penta-O, where we use circuit parameters to represent the circuits for $e^{-i\gamma _l H_{\rm C}}$ and $e^{-i\theta _l H_{\rm M}}$. In the diagram, gray, blue, and red denote inherited circuits, changeable circuits, and measurements, respectively. The number of trials required in each step is indicated by the number on top of the measurement. (b) Explanation of MaxCut. Given a graph, the vertices are partitioned into two groups (red and blue), and we ``cut" the edges connecting the two groups. MaxCut seeks the bipartition that maximizes the number of cut edges. By assigning spin values to represent the two groups (for example, $\uparrow$ as blue and $\downarrow$ as red), MaxCut can be reformulated as a field-free Ising model, where the graph edges define the interaction between spins. The optimal solution to the MaxCut corresponds to the ground state of this spin system.}
\end{figure*}

\textit{Level-wise construction.} --- By substituting binary variable (bit) $x \in \{0, 1\}$ with spin variable $z=1-2x$, the optimal solution of a generic QUBO problem can be encoded as the ground state of a cost Hamiltonian up to an unimportant constant~\cite{lucas2014ising,glover2022quantum,abbas2024challenges}:
\begin{equation}
    H_{\rm C} = \sum _{i<j} w_{ij} Z_i Z_j + \sum _{i=1}^{N} w_{ii} Z_i \label{eq:qubo},
\end{equation}
where $N$ is the number of qubits, $i, j \in \{1,2,...,N\}$, $Z_i$ is the Pauli $Z$ operator for the $i$-th qubit, and $w_{ij} \in \mathbb{R}$ is the reduced weight. QAOA samples the ground state of $H_{\rm C}$ with a parametric quantum circuit of the form
\begin{equation}
    |\psi _p \rangle = [\prod _{l=1}^{p} e^{-i\gamma _l H_{\rm M}} e^{-i\theta _l H_{\rm C}}] |+ \rangle,
\end{equation}
where $p$ is referred to as the \textit{level} of QAOA, $\gamma _l , \theta _l \in \mathbb{R}$ are \textit{circuit parameters} of the $l$-th level, $H_{\rm M}= - \bigotimes _{i=1}^{N} X_i$ is the mixer Hamiltonian consists of $N$ Pauli $X$ operators, and $| + \rangle = \bigotimes _{i=1}^{N} R_Y(\frac{\pi}{2}) |0\rangle ^{\otimes N}$ denotes the initial state for simplicity. As $\gamma _l , \theta _l \to 0$ and $p \to +\infty$, QAOA becomes a Trotterized approximation of quantum annealing~\cite{hauke2020perspectives,PhysRevE.58.5355,farhi2000quantum}. However, for finite $p$, the circuit parameters are typically optimized by minimizing the objective function
\begin{equation}
    J _p = \langle \psi _p | H_{\rm C} | \psi _p \rangle, \label{eq:jp}
\end{equation}
which represents the energy expectation. Throughout the text, we refer to a single attempt to estimate $J_p$ as one trial. In experiments, this process is highly resource-intensive and constitutes the primary source of sampling overhead, as $J_p$ is constructed from the quasi-probabilities obtained by repeatedly preparing and measuring $|\psi _p \rangle$.

We extend the established works that derive the analytical form of $J_p$ for $p=1$~\cite{PhysRevA.97.022304,PhysRevLett.125.260505,Bravyi2022hybridquantum,Farhi2022quantumapproximate,ozaeta2022expectation}, and demonstrate that $J_p$ is a simple trigonometric function of $\theta _p$ (see Section S1 in the supplemental materials for the proof~\cite{sppm}):
\begin{equation}
    J _p = A_p \sin (4 \theta _{p} + \phi_p ) + A_p^{\prime} \sin (2 \theta _{p} + \phi_p ^{\prime}) + C_p, \label{eq:general_sin}
\end{equation}
where $A_p$, $A_p^{\prime}$, $\phi_p$, $\phi_p^{\prime}$, and $C_p$ are functions of $\gamma _p$ and the circuit parameters of the preceding levels, but, surprisingly, are independent of $\theta _p$. Calculating these coefficients on classical computers is exponentially difficult~\cite{Farhi2022quantumapproximate,farhi2020see1,farhi2020see2}, whereas quantum computers, leveraging their sampling advantage, require at most five trials.

Our analytical results naturally give rise to a level-wise parameter-setting strategy, as illustrated in Fig.~\ref{fig:overview}(a). For simplicity, we assume $\gamma _l = \gamma _0$ for all levels and defer the discussion on the choice of $\gamma _0$ to a later part. The algorithm begins with $p=1$, where we search for the optimal $\theta _1^*$ that minimizes $J_1$. Since $J_p$ has a period of $\pi$, it is straight-forward to derive the $\theta _p$-independent coefficients with 5 trials, each using one of the probe angles $\theta _x = \frac{k}{6}\pi \space (k=1,2,3,4,5)$. Then $\theta _1^*$ is derived based on Eq.~\ref{eq:general_sin}. Subsequently, the double-level QAOA is constructed by inheriting the $p=1$ circuits with parameters $(\gamma _0, \theta _1^*)$, and $\theta _2^*$ is derived following the same procedure. Repeating this step $p$ times generates a $p$-level QAOA with optimized circuit parameters, which is ultimately used to approximate the ground state of $H_{\rm C}$. 

Due to its level-wise structure, the time complexity of preparing all quantum circuits is a sum of an arithmetic sequence, resulting in a quadratic scaling $\mathcal{O}(p^2)$ and a linearly scaled sampling overhead $(5p + 1)M$, where $M$ is the number of repetition for one trial. Notably, our proof also enforces a $\pi/2$ period when $w_{ii}=0$, which consequently implies that both $A_p^{\prime}$ and $\phi _p ^{\prime}$ vanish. In consequence, the number of trials required at each step is reduced to 3 when $w_{ii}=0$, with probe angles $\theta _x = \frac{k}{8}\pi \space (k=1,2,3)$ used in this case. Finally, since $J_p = J_{p-1}$ if $\theta _p =0$, there always exists a $\theta _p$ such that $J_p \leq J_{p-1}$, guaranteeing non-decreasing performance. As estimating the objective function at each step generally requires five trials, we name the parameter-setting strategy as ``\textbf{Penta-O}".

\textit{Benchmark results} --- Our benchmark results start with MaxCut, a quintessential benchmark for evaluating QAOA. Given an undirected graph $\mathcal{G}$, MaxCut aims to find an optimal bipartition of vertices that maximizes the number of edges connecting the two groups. Up to a constant, the cost Hamiltonian for MaxCut is
\begin{equation}
    H_{\rm C} = \sum _{(i,j) \in E} w_{ij} Z_i Z_j, \label{eq:maxcut_hc}
\end{equation}
where the index pairs $(i,j)$ span the edge set $E$ of $\mathcal{G}$, and $w_{ij}$ represents edge weights. As illustrated in Fig.~\ref{fig:overview}(b), MaxCut is fundamentally equivalent to a field-free Ising model. The algorithm's performance on MaxCut is evaluated using the approximation ratio~\cite{PhysRevX.10.021067}:
\begin{equation}
    r = \frac{W - J_p}{W - J_{\rm min}},
\end{equation}
where $W = \sum _{i=1}^N w_{ij}$ is the total weight of all edges, and $J_{\rm min}$ is the ground state energy, which is obtained via brute-force search in this work. $r=1$ corresponds to the optimal solution. Notably, it is \textit{NP}-hard to design an algorithm that guarantees a performance better than $r_{\rm g}^* = 16/17$ across generic graphs~\cite{berman1999some}. The best classical algorithm, developed by Goemans and Williamson~\cite{goemans1995improved}, provides a lower bound of $r_{\rm g} \approx 0.8786$, corresponding to a probability of 0.25 to sample the optimal solution.

\begin{figure*}
\includegraphics[width=1.0\textwidth]{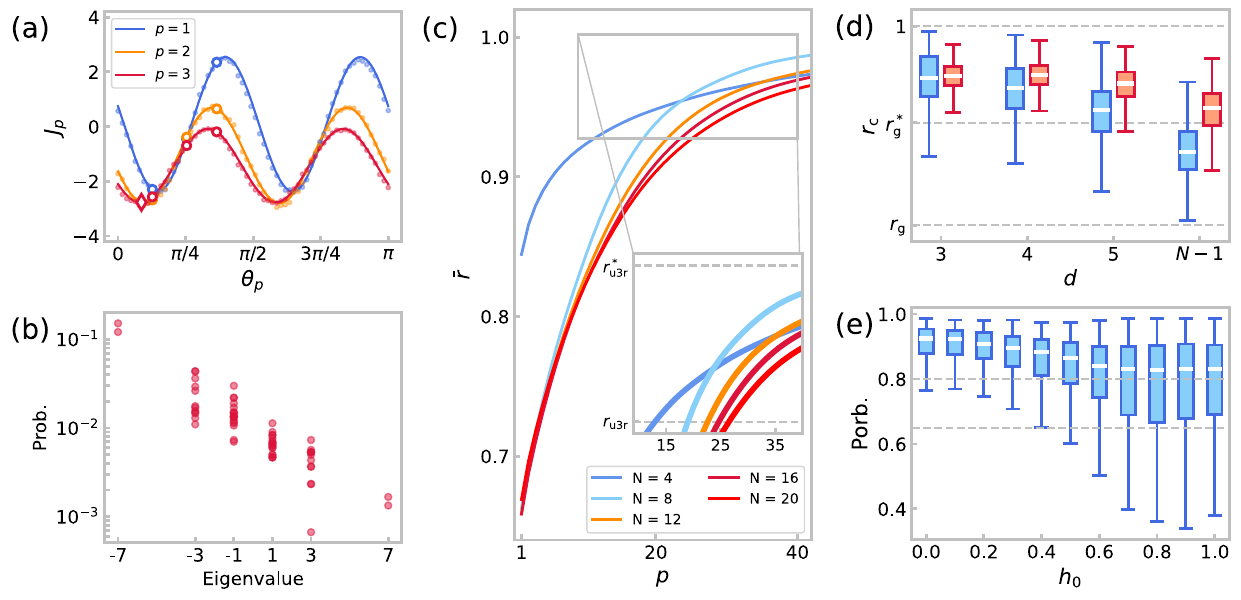}
\caption{\label{fig:2} (a) Triple-level QAOA experiment results with Penta-O where $\gamma _0 = 0.2$. Each data point corresponds to one trial with $M=3000$. (b) Probabilities of solutions obtained in the final trial, which is also conducted with 3000 repetitions. (c) The average approximate ratio obtained with $\gamma _0 = 0.075$ as a function of $p$. All nonisomorphic graphs are accessible via the open-access database \textit{The House of Graphs}~\cite{houseofgraphs}. The anomalous behavior observed for $N=4$ is attributed to the fact that such a graph is actually a complete graph. (d) Box plot of approximate ratio at convergence $r_{\rm c}$ for different degree $d$ whose weights are sampled from P-d (blue) and N-d (red). We define $r$ as converged if the improvement achieved by adding a new level is less than $5/1000$. (e) Box plot of the probability of sampling the low-energy states at different fields $h_0$. Horizontal dash lines are plotted to visualize the critical chance of 0.65 and 0.8. The box plot uses boxes to indicate interquartile range (IQR) between the 25th and 75th percentiles, with a horizontal line (white) to denote the median. Boundaries of whiskers extending outside the boxes are based on the 1.5 IQR value.}
\end{figure*}

As a proof of principle, we conduct experiments on a superconducting quantum processor to validate our method, which are performed on the same platform as our previous works~\cite{PhysRevLett.133.170601}. Specifically, we consider the MaxCut problem on an unweighted $2 \times 3$ grid graph, which aligns with the native topological connectivity of the chip. Fig.~\ref{fig:2}(a) illustrates the parameter landscape of $p=1,2,3$. The experiments construct the circuits level-by-level with three trials at each step, with corresponding energy expectation marked with open circles. The derived $J_p$ plotted as the solid lines are in good agreement with the experiment results represented by scatter points. The lowest energy expectation monotonically decreases as more levels are added, reaching optimal performance at $p=3$. Experiments with larger $p$ show negligible performance improvement, likely caused by increasing qubit dephasing. Fig.~\ref{fig:2}(b) illustrates the probability of solutions grouped by their eigenvalues obtained in the final step, corresponding to the experiment marked in a diamond in Fig.~\ref{fig:2}(a). The ground state probability $\approx 0.2733$ is comparable to the classical algorithm, validating our approach.


To further demonstrate the performance of our method, we perform noiseless simulations using \textit{Qiskit}~\cite{javadi2024quantum} on unweighted 3-regular graphs (u3r). Characterized by a fixed degree $d=3$, meaning that each vertex is connected to exactly three edges, this class of graph is crucial in complexity theory~\cite{alimonti2000some}. When restricted to u3r, the performance lower bound of classical algorithm is improved to $r_{\rm u3r} \approx 0.9326$~\cite{halperin2004max}, and the \textit{NP}-hard threshold is raised to $r_{\rm u3r}^* = 331/332$~\cite{berman1999some}. We conduct a comprehensive analysis of all nonisomorphic graphs with $N \leq 14$ and study randomly generated replicas for $14 < N \leq 20$. The approximate ratio averaged among the replicas are collected in Fig.~\ref{fig:2}(c). Our method archives a ratio above the classical lower bound with $p \approx 30$ for average cases, and $p \approx 40$ for the worst case (see Section S2 in the supplemental materials~\cite{sppm}). This marks a substantial improvement over FALQON~\cite{PhysRevLett.129.250502,PhysRevA.106.062414}, the most prominent outer-loop-free method to date, which requires $p \approx 500$ for the worst case.

We further explore its effectiveness on weighted graphs, which are challenging due to their intricate parameter landscapes~\cite{shaydulin2023parameter}. We focus on weighted $d$-regular graph (w$d$r), where the edge weights are sampled from two representative distributions: (1) the non-negative Poisson distribution (P-d) with a mean $\lambda = 1$, and (2) the Normal distribution (N-d), which includes negative weights. 
Fig.~\ref{fig:2}(d) summarizes the algorithmic performance for $N=14$, with detailed results provided in Section S2 in the supplemental materials~\cite{sppm}. The approximate ratio at convergence, $r_{\rm c}$, exhibits comparable performance across different degrees and weight distributions, where more than 75\% replicas exceeding the classical lower bound $r_{\rm g}$. Notably, it shows remarkable robustness on graphs with negative weights (N-d), where the classical algorithms often struggle with the divergence in the ratio between the semidefinite relaxation and the integral optimum~\cite{charikar2004maximizing}. Our results indicate that QAOA enhanced by Penta-O holds significant promise for densely connected graphs or those with negative weights.

Finally, we consider $w_{ii} \neq 0$, which corresponds to Ising model with longitudinal field. Specifically, we analyze the Sherrington-Kirkpatrick (SK) model~\cite{panchenko2013sherrington,talagrand2010mean}, a cornerstone of spin glass theory. It describes a classical spin system with all-to-all couplings among $N$ spins, where the coupling strengths are drawn from a random distribution with mean 0 and variance 1. The cost Hamiltonian maintains the same structure as Eq.~\ref{eq:qubo}. We assume a homogeneous field ($w_{ii} = h_0$) and $w_{ij} = \pm 1$ are chosen with equal probability. We examine systems under varying external fields $h_0$, ranging from $0$ to $1$ in increments of $0.1$. For each $h_0$, 500 replicas with different coupling configurations are generated, resulting in a total of $11 \times 500$ instances. We analyze the success probability of measuring low-energy states with $\gamma _0 = 0.05$, defined as states with a normalized eigen-energy $ < 0.1$ (see Section S2 in the supplemental materials for details~\cite{sppm}). As depicted in Fig.~\ref{fig:2}(e), for half the replicas, the probability of measuring low-energy states exceeds 0.8. Approximately $75\%$ of the replicas exhibit a probability above 0.65, but even in the worst-case scenario, the probability remains above 0.3. These findings indicate that our method effectively simulates low-energy states of spin systems.

\textit{Discussion and outlook} --- We present Penta-O as a novel paradigm to enhance QAOA performance on QUBO problems, validated through both experiment and simulation. The key feature of Penta-O is its outer-loop-free algorithm for parameter-setting, which scales quadratically with the QAOA level $p$ and requires no more than $5p$ trials to derive all circuit parameters. Benchmarking on MaxCut and the SK model with longitudinal field demonstrates that the circuit parameters generated by Penta-O provide consistently non-decreasing, near-optimal performance across a diverse range of problem instances. To the best of our knowledge, Penta-O is the most efficient parameter-setting strategy that combines non-decreasing performance with bounded computational complexity.

Throughout the text, we adopt a fixed $\gamma _0$ for all levels, a novel approach inspired by control theory. Denoting $\gamma _0 = \Delta t$ and rewriting $\theta _l = u_l \cdot \Delta t$, the alternating application of $H_{\rm C}$ and $H_{\rm M}$ can be interpreted as a discretization of the evolution of $H_{\rm C} + u(t) H_{\rm M}$, where $H_{\rm M}$ represents the control field with a time-dependent amplitude $u(t)$~\cite{PhysRevLett.129.250502,PhysRevA.106.062414}. From this perspective, Penta-O serves as a near-optimal control for cooling down the system when $\Delta t = \gamma _0 \ll 1$. The error of the discretization approximation is proportion to $[H_{\rm C}, H_{\rm M}] \Delta t$, which accounts for the performance decline observed in Fig.~\ref{fig:2}(d) and Fig.~\ref{fig:2}(e). This approach holds significant potential for refinement, as the constant $\gamma _0$ contrasts with the monotonically increasing optima of $\gamma _l$ proposed in Ref.~\cite{farhi2014quantum}. However, determining the optimal value for each level remains an open question for future research, given the highly non-linear relationship between $\gamma _l$ and $w_{ij}$. Potential method could involve machine learning or a feedback strategy~\cite{brady2024focqs}.

The sampling advantage of quantum computers, which allows for the estimation of $J_p$ in polynomial time, offers valuable insights into achieving practical quantum advantage. The time cost of a Penta-O-enhanced $p$-level QAOA is given by
\begin{equation}
    T_{\rm q} = [\frac{5p(p+1)}{2} + p] \cdot M t_0,
\end{equation}
where $t_0 \propto N $ represents the time required to prepare a single-level QAOA and $M t_0$ is the time cost of one trial. We assume that $M = 10^3$ repetitions per trial are sufficient for complete graphs when $N < 1000$~\cite{lotshaw2022scaling}. For a MaxCut problem of non-trivial size, the time-to-solution for a classical solver scales exponentially with $N$~\cite{guerreschi2019qaoa}. Under practical assumptions, we propose that the quantum-classical crossover in time-to-solution is anticipated when $N \approx 500$ (see Section S3 in the supplemental materials for details~\cite{sppm}). Specifically, if the $p$-$N$ scaling outperforms linear growth, quantum computers could solve MaxCut for complete graphs within an hour, whereas classical solvers might require days or even decades. 

As the potential of deep-depth QAOA remains largely unexplored due to the growing computational demands of parameter optimization, we believe this study will inspire further exploration of variational algorithms and quantum annealing. A promising direction for future research is to investigate the generalized Ising model defined on a hypergraph, which enables $k$-local spin coupling $\prod _{i=1}^k Z_i$. Other intriguing avenues include investigating phenomena associated with deep-depth circuits, such as nonadiabatic mechanisms~\cite{PhysRevX.10.021067}, pseudo-Boltzmann distribution~\cite{PhysRevLett.130.050601,PhysRevA.108.042411}, locality of states~\cite{PhysRevLett.125.260505,hastings2019classical}, and the critical $p$-$N$ scaling law~\cite{PhysRevA.108.042411,PhysRevA.106.042438,PRXQuantum.4.010309}.

\begin{acknowledgments}
This work was supported by the Innovation Program for Quantum Science and Technology (2021ZD0301703), the Shenzhen-Hong Kong Cooperation Zone for Technology and Innovation (HZQB-KCZYB-2020050), and Guangdong Basic and Applied Basic Research Foundation (2024A1515011714, 2022A1515110615).
\end{acknowledgments}

\bibliography{main_ref}

\end{document}



\title{Supplemental materials for: Improving Quantum Optimization to Achieve Quadratic Time Complexity}

\maketitle

\section{Proof of the algorithm base ground}
In the main text, we assert that the expectation value of a $p$-level QAOA, $J_p$, is a simple trigonometric function of $\theta _p$. Specifically, we express it as:
\begin{equation}
    J _p = A_p \sin (4 \theta _{p} + \phi_p ) + A_p^{\prime} \sin (2 \theta _{p} + \phi_p ^{\prime}) + C_p. \label{eqs:general_sin}
\end{equation}
Here, we assume that all circuit parameters for the preceding $p-1$ levels are fixed, making all variables except $\theta _p$ constants. This equation forms the foundation of our algorithm, and the proof is provided below.

\bigskip

Given an $N$-spin Ising model with external fields where the coupling is defined on an undirected graph $\mathcal{G}$, the cost Hamiltonian is
\begin{equation}
    H_{\rm C} = \sum _{i\in E} w_{ij} Z_i Z_j + \sum _{i=1}^{N} w_{ii} Z_i \nonumber,
\end{equation}
where $E$ is the edge set of $\mathcal{G}$. Following Eq.~4 in the main text, the state prepared by a $p$-level QAOA is
\begin{equation}
    |\psi _p \rangle = [\prod _{l=1}^{p} e^{-i\gamma _l H_{\rm M}} e^{-i\theta _l H_{\rm C}}] |+ \rangle \nonumber.
\end{equation}
Denote the evolution of the cost and mixer Hamiltonian as
\begin{equation}
    U_{\rm C}(l) = e^{-i \gamma _l H_{\rm C}},  \nonumber \\
    U_{\rm M}(l) = e^{-i \theta _l H_{\rm M}},  \nonumber
\end{equation}
the alternating application of $U_{\rm C}(l)$ and $U_{\rm M}(l)$ yields a $p$-level energy expectation:
\begin{equation}
    J_ {p} = \langle + | U_{\rm C}^{\dagger}(1) U_{\rm M}^{\dagger}(1) ... U_{\rm C}^{\dagger}(p) \cdot \underbrace{U_{\rm M}^{\dagger}(p) H_{\rm C} U_{\rm M}(p)}_{\mathcal{J}}  \cdot U_{\rm C}(p) ... U_{\rm M}(1) U_{\rm C}(1) | + \rangle.  \nonumber
\end{equation}
The operators indicated by the under-brace is:
\begin{align}
    \mathcal{J} =& \sum _{(i,j) \in E} [ U_{\rm M}^{\dagger}(p) \cdot w_{ij} Z_i Z_j \cdot U_{\rm M}(p) ] \nonumber \\
                 &+ \sum _{i = 1}^N  [ U_{\rm M}^{\dagger}(p) \cdot w_{ii} Z_i \cdot U_{\rm M}(p) ] \nonumber \\
                =& \sum _{(i,j) \in E} \mathcal{J}_{ij} + \sum _{i = 1}^N \mathcal{J}_{ii} . \nonumber
\end{align}

Notice the identity~\cite{ozaeta2022expectation}:
\begin{equation}
    U_{\rm M}^{\dagger}(p) Z_i U_{\rm M}(p) = \cos 2 \theta_p \cdot Z_i + \sin 2 \theta _p \cdot Y_i.
\end{equation}
For notation simplicity, let $c_2 = \cos 2 \theta _p$, $s_2 = \sin 2 \theta _p$. $\mathcal{J}_{ii}$ and $\mathcal{J}_{ij}$ can be simplified as following:
\begin{align}
    \mathcal{J}_{ij} &= w_{ij} U_{\rm M}^{\dagger}(p) Z_i Z_j U_{\rm M}(p) \nonumber
    \\
    &= w_{ij} \cdot U_{\rm M}^{\dagger}(p) Z_i U_{\rm M}(p) \cdot U_{\rm M}^{\dagger}(p) Z_j U_{\rm M}(p)  \nonumber \\
    &= w_{ij} (c_2 Z_i + s_2 Y_i)(c_2 Z_j + s_2 Y_j), \label{eqs:jij}
\end{align}
and
\begin{align}
    \mathcal{J}_{ii} &= w_{ii} U_{\rm M}^{\dagger}(p) Z_i U_{\rm M}(p) \nonumber
    \\
    &= c_2 Z_i + s_2 Y_i. \label{eqs:jii}
\end{align}
Notably, the mixer parameter $\theta _p$ has been separated from all Pauli operators and becomes as scalar terms, which implies that $\mathcal{J}$ can be written in the form of a trigonometric function:
\begin{equation}
        \frac{1}{w_{ij}} \langle \psi _{p-1} | U_{\rm C}^{\dagger}(p) \cdot \mathcal{J}_{ij} \cdot U_{\rm C}(p) | \psi _{p-1} \rangle 
       = c_2^2 \langle Z_i Z_j \rangle _{\gamma _p,p-1}  + s_2^2 \langle Y_i Y_j \rangle _{\gamma _p,p-1} + c_2 s_2 \langle Z_i Y_j + Y_i Z_j \rangle _{\gamma _p, p-1}, \label{eqs:4theta}
\end{equation}
\begin{equation}
        \frac{1}{w_{ii}} \langle \psi _{p-1} | U_{\rm C}^{\dagger}(p) \cdot \mathcal{J}_{ii} \cdot U_{\rm C}(p) | \psi _{p-1} \rangle 
       = c_2 \langle Z_i \rangle  _{\gamma _p, p-1}  + s_2 \langle Y_i \rangle _{\gamma _p, p-1}, \label{eqs:2theta}
\end{equation}
where $\langle \cdot \rangle _{\gamma _p,p-1}$ indicates the expectation is subjected to the state $U_{\rm C}(p)|\psi _{p-1} \rangle$. Summing over the edges and vertices leads to the energy expectation of the $p$-level QAOA:
\begin{align}
    J_p = c_2^2 \sum_{(i,j) \in E} w_{ij} \langle Z_i Z_j \rangle _{\gamma _p,p-1}  &+ s_2^2 \sum_{(i,j) \in E} w_{ij} \langle Y_i Y_j \rangle _{\gamma _p,p-1} \nonumber \\
    &+ c_2 s_2 \sum_{(i,j) \in E} w_{ij} \langle Z_i Y_j + Y_i Z_j \rangle _{\gamma _p, p-1} \nonumber \\
         &+ c_2 \sum_{i=1}^N w_{ii} \langle Z_i \rangle _{\gamma _p, p-1}  + s_2 \sum_{i=1}^N w_{ii} \langle Y_i \rangle _{\gamma _p, p-1}. \label{eqs:sin_general}
\end{align}
As $c_2^2$, $s_2^2$, and $c_2 s_2$ share the same period of $\pi/2$, and $c_2$ and $s_2$ share the same period of $\pi$, Eq.~\ref{eqs:sin_general} can be simplified by introducing extra phases and offset, which in the end leads to the form as Eq.~\ref{eqs:general_sin}. In particular, we have the corresponding coefficients
\begin{equation}
    \left\{
    \begin{aligned}
        A_p &= \frac{1}{2} \sqrt{(O_{ZZ} - O_{YY})^2 + O_{ZY}^2},  \\
        \phi _p &= \arctan{\frac{O_{ZZ} - O_{YY}}{O_{ZY}}},  \\
        A _p^{\prime} &= \sqrt{O_{Z}^2 + O_{Y}^2},  \\
        \phi _p^{\prime} &= \arctan{\frac{O_{Z}}{O_{Y}}},  \\
        C_p &= \frac{O_{ZZ} + O_{YY}}{2},
    \end{aligned}
    \right. \label{eqs:exp_params}
\end{equation}
where we denote expectation of the observables as
\begin{equation}
    \left\{
    \begin{aligned}
        O_{ZZ} &= \sum_{(i,j) \in E} w_{ij} \langle Z_i Z_j \rangle _{\gamma _p,p-1},  \\
        O_{YY} &= \sum_{(i,j) \in E} w_{ij} \langle Y_i Y_j \rangle _{\gamma _p,p-1},  \\
        O_{ZY} &= \sum_{(i,j) \in E} w_{ij} \langle Z_i Y_j + Y_i Z_j \rangle _{\gamma _p, p-1},  \\
        O_{Z} &= \sum_{i=1}^N w_{ii} \langle Z_i \rangle _{\gamma _p, p-1},  \\
        O_{Y} &= \sum_{i=1}^N w_{ii} \langle Y_i \rangle _{\gamma _p, p-1}.
    \end{aligned}
    \right. \label{eqs:obs}
\end{equation}
The values are independent of $\theta _p$, thereby prove Eq.~\ref{eqs:general_sin}.

\bigskip

When $w_{ii} = 0$, we have $O_{Z}=O_{Y}=0$, and the term with a period of $\pi$ in Eq.~\ref{eqs:general_sin} vanishes, consistent with the statement mentioned in the main text. Although the coefficients can be derived from the measurement results of these observables, approximately $d$ rounds of measurement are required because the observables in Eq.~\ref{eqs:obs} do not commute, where $d$ is the degree of the graph $\mathcal{G}$. Consequently, we estimate $J_p$ at different probe angles, which generally reduces the sampling overhead. We also emphasize that the it is exponentially hard to calculate the coefficients for $p > 1$ with a classical computer. By leveraging the sampling advantage of quantum computers, we may obtain the full knowledge of $J_p$ with polynomial time complexity.

The expectation of the observables are generally non-linear functions of $w_{ij}$, except for $O_{ZZ}$ and $O_{Z}$ as $[Z_i Z_j, Z_i]=0$. Since $U_{\rm C}(p)$ incorporates the edge weights and $\gamma _p$, observables that do not commute with $U_{\rm C}(p)$ produce scalar terms in the form of $\prod \cos (2 w_{ij} \gamma _p)$, where the range of the product depends on the graph connectivity. Consequently, deriving the analytical form and determining the optimal $\gamma _p$ for generic graphs poses a significant challenge. For more detailed discussions, readers are referred to the appendix in Ref.~\cite{ozaeta2022expectation} and the supplemental materials in Ref.~\cite{PhysRevLett.125.260505}.

\begin{figure*}
\includegraphics[width=1.0\textwidth]{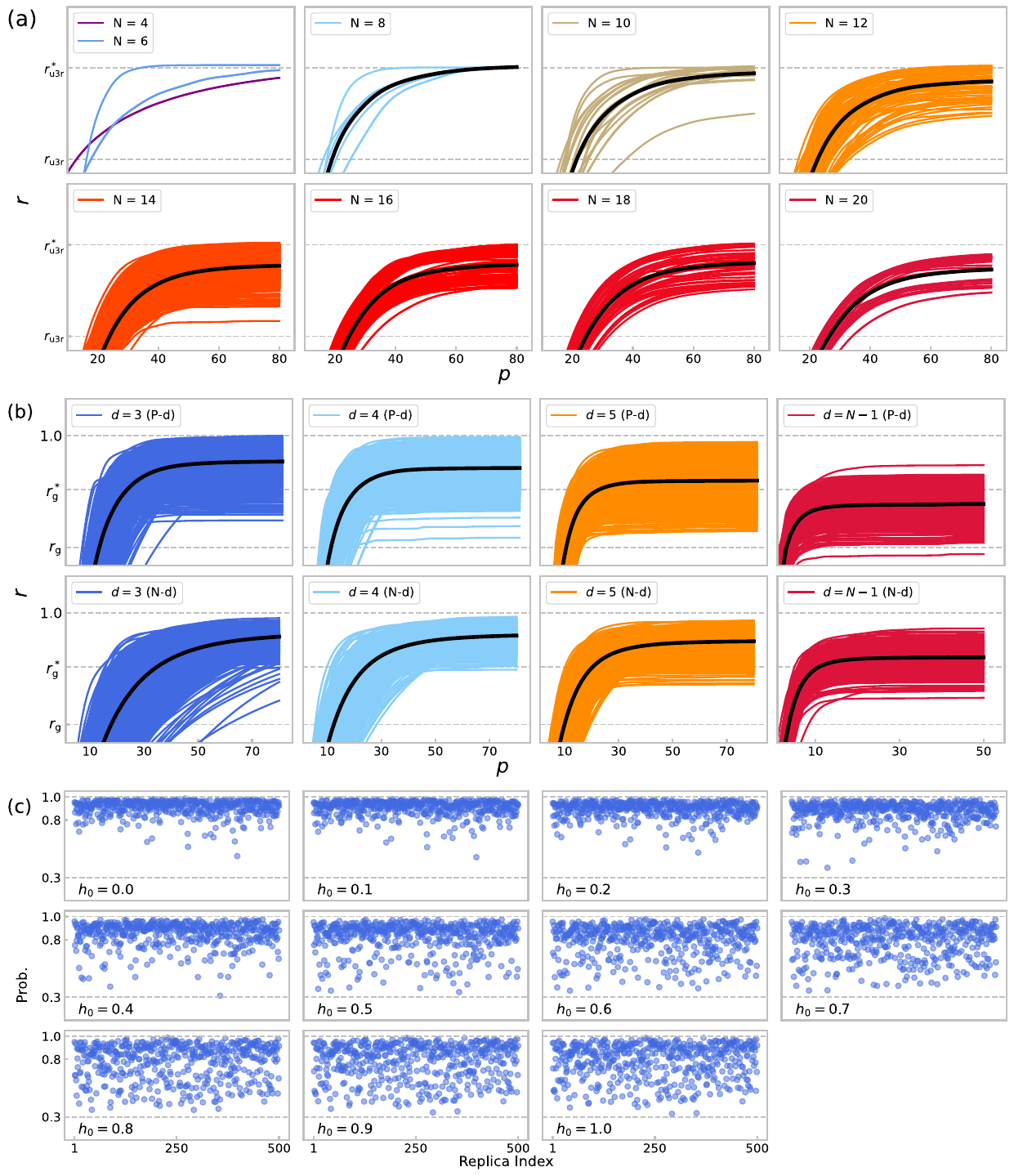}
\caption{\label{figs:simu} The approximate ratio as a function of QAOA level on u3r with different size $N$ (a) and weighted graphs of $N=14$ (b). The colored lines are the approximate ratio as a function of $p$ for individual replicas. Solid lines in black are the ratio averaged among all replicas of the same size. Probability of sampling the low-energy states for a SK model with a uniform external field $h_0$ is given in (c).}
\end{figure*}

\section{Numerical results in details}
The results shown in Fig.~2(c) in the main text are limited to $p \leq 40$. In Fig.~\ref{figs:simu}(a), we provide the full simulation results for $p=80$ across all instances. Graphs of the same size are grouped in the same diagram except for $N=4$ and $N=6$, as there are only three nonisomorphic graphs in total. The number of nonisomorphic graphs with sizes $N=8,10,12,14$ are $5,19,85,509$, respectively. Due to the increasing computational demands in larger systems, we randomly selected $100,50,25$ nonisomorphic replicas for $N=16,18,20$, respectively. All instances are accessible via the open-access database \textit{The House of Graphs}~\cite{houseofgraphs}. The solid black line represents the average approximation ratio, consistent with the data shown in Fig.~2(c). We believe these instances adequately capture the general trend for larger system sizes. In particular, the results shown here illustrate the worst-case scenario, where a 40-level QAOA is necessary for the instance with $N=10$ to archive $r > r_{\rm u3r}$. Additionally, we observe no indications of surpassing $r_{\rm u3r}^*$ with a 80-level QAOA, except in rare cases where $N < 20$.

Fig.~\ref{figs:simu}(b) shows all replicas used to generate Fig.~2(d) and Fig.~2(e) in the main text. The first row corresponds to replicas with weights sampled from the Poisson distribution, while the second row depicts those sampled from the Normal distribution. The columns represent w3r, w4r, w5r, and the complete graph, respectively. The average across all replicas of the same graph type is also indicated by a solid black line. The presented results, excluding complete graphs ($d=N-1$), are averaged over $100 \times 10$ replicas, consisting of 100 nonisomorphic unweighted graphs, each combined with 10 sets of random weights. For complete graphs, 300 replicas were found sufficient to achieve stable averages in the results. To constrain the magnitude of $\gamma _0$, the cost Hamiltonian $H_{\rm C}$ has been normalized by the maximum $|w_{ij}|$. For graphs with negative weights, the definition of $W_E$ has been adjusted to $W = \sum _{(i,j) \in E} |w_{ij}|$, which avoids possible singularity and does not affect overall trend. Given the similar convergence trends observed among different replicas, we believe these instances effectively capture the algorithm's performance.

Fig.~2(e) of the main text illustrates the probability of sampling the low-energy states for SK model with different $h_0$. We normalize the eigenvalues of $H_{\rm C}$ into the $[0,1]$ interval, and the energy expectation $J_p$ is reduced to the normalized accordingly. We define the low-energy states as those whose normalized eigenvalues $< 0.1$. Fig.~\ref{figs:simu}(c) presents the probability of low-energy states of all replicas investigated. Observe that there is a chance larger than 0.3 of sampling the low-energy states for all replicas.

\begin{figure}
\includegraphics[width=0.9\textwidth]{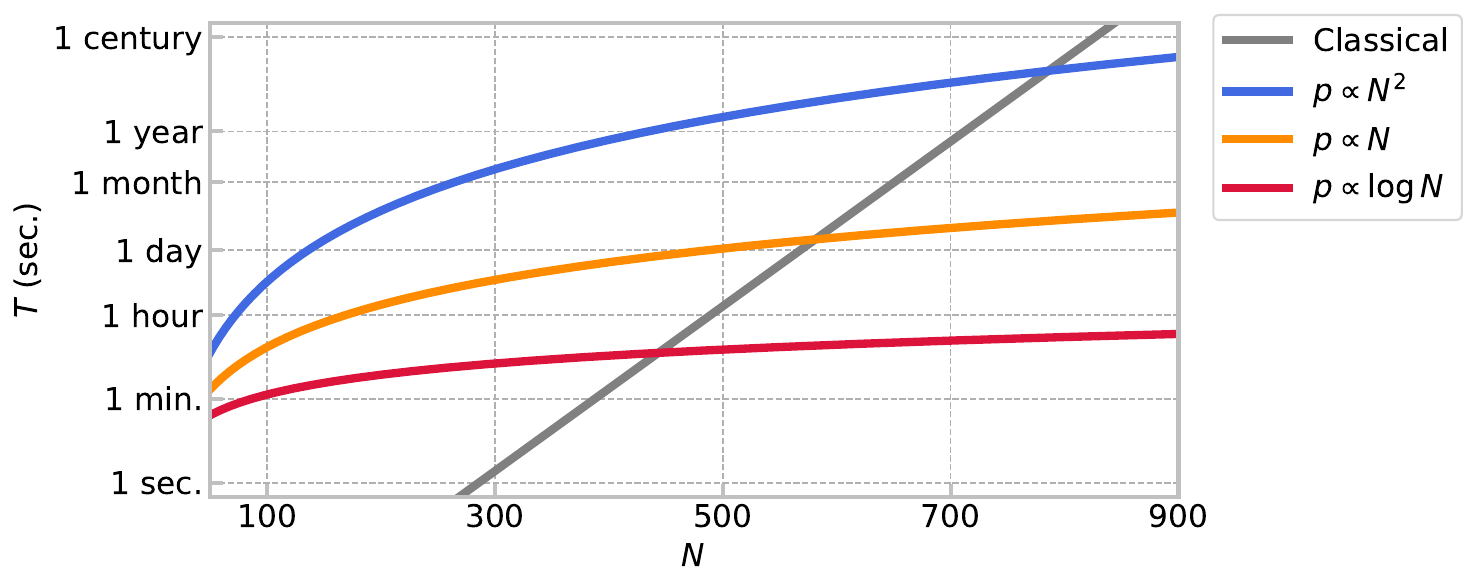}
\caption{\label{figs:adv} Time-to-solution of a quantum and classical solver as functions of problem size evaluated for MaxCut of complete graphs under general assumptions.}
\end{figure}

\section{Analyze the quantum advantage}
Since the initial state and the evolution of mixer Hamiltonian $e^{-i\theta _l H_{\rm M}}$ can be efficiently prepared with $N$ concurrent single qubit gates, the time cost of a $p$-level QAOA is primarily determined by the $p$ repetitions of the circuit $e^{-i\gamma _l H_{\rm C}}$. Let $t_0$ denote the time required to prepare a single-level QAOA. The time cost for a $l$-level QAOA is approximately $l t_0$. To estimate $J_l$, the same circuit must be prepared and measured $M$ times, resulting in a total time cost of $M l t_0$. Penta-O facilitates $p$ steps to obtain parameters for a $p$-level QAOA, where an $l$-level QAOA is repeated for 5 trials. This process results in a total time cost of
\begin{align}
    T_{\rm q} &= \sum _{l=1}^p 5M \cdot l \cdot t_0 + M p t_0 \nonumber \\
    &= [\frac{5p(p+1)}{2} + p] \cdot M t_0,
\end{align}
where we include the time cost of the final step $M p t_0$, which is used to approximate the ground state with the optimized circuit parameters.

When the problem is defined on a graph, the set of two-qubit gates that can be applied simultaneously is determined by solving a corresponding edge-coloring problem. An efficient classical algorithm~\cite{misra1992constructive} provides a solution with at most $d+1$ layers of two-qubit gates, where $d$ is the degree of the graph. In the worst-case scenario, represented by a complete graph, $N$ layers of concurrent gates are required. On a typical superconducting quantum processor, the duration of a two-qubit gate, $\tau$, ranges from 20~ns to 300~ns depending on the implementation method. Here we assume $\tau \sim 500$~ns, which includes the implementation of constant-depth long-range two-qubit gates~\cite{baumer2024measurement}. This corresponds to an estimated $t_0 \sim N/2$~$\mu$s. Following the scaling law proposed in Ref.~\cite{lotshaw2022scaling}, we assume that $M = 10^3$ repetitions per trial are sufficient for complete graphs when $N < 1000$. To date, the relationship between the QAOA level $p$ and the problem size $N$ remains uncertain, except for the lower bound $p > \log N$~\cite{farhi2020see1,farhi2020see2,PhysRevLett.125.260505,PRXQuantum.4.010309}. Given the clear saturation of convergence observed in the cases we studied, we propose three potential $p$-$N$ scaling laws: $p_1(N) = \alpha _1 N ^ 2$, $p_2(N) = \alpha _2 N$, and $p_3(N) = \alpha _3 \log N$. The constants $\alpha _i \space (i=1,2,3)$ are determined such that $p_i (20) = 30$, which aims to cover the worst case shown in Fig.~\ref{figs:simu}(b). Based on these assumptions, we conduct a symbolic comparison of the time resources required by quantum and classical solvers.

For a MaxCut problem of non-trivial size, the time-to-solution (measured in seconds) for a classical solver scales exponentially with $N$, approximated by~\cite{guerreschi2019qaoa}
\begin{equation}
    T_{\rm c} = 10^{-5} e^{0.04029 N}.
\end{equation}
The small exponent highlights the capability of the highly advanced classical solver, which solves a 300-qubit problem within few seconds. To demonstrate quantum advantages in problems of interest, quantum computers must surpass this scale significantly. Fig.~\ref{figs:adv} illustrates the time-to-solution, $T$, as a function of problem size $N$. Our results suggest that QAOA does not provide a practical advantage if $p \propto N^2$, as the crossover occurs when $T _{\rm q}$ exceeding one year. Quantum advantage begins to emerge for $p \propto N$ at a problem size of approximately $N \approx 600$, which still requires few days. Significant quantum advantage is observed only when $p \propto \log N$, where a problem size of $N = 900$ is solved within an hour, whereas the classical solver requires more than a century. Under these conditions, we also anticipate that quantum advantage begins to emerge at approximately $N \approx 500$. The assumption ignores the cost brought by quantum error correction because there lacks practical estimation for the time cost of logical two-qubit gates. If the order of magnitude of the time cost a logical operation is 10~$\mu$s, it is only possible to observe practical quantum advantage if $p \propto \log N$, whereas the quantum-classical crossover of $p \propto N$ is increased to more than one year.

\bibliography{main_supple_ref}